\documentstyle[aps,preprint,epsf]{revtex}
\begin{document}
\draft
\title{Spectral scrambling in Coulomb-blockade quantum dots}

\author{ Y. Alhassid$^1$ and Yuval Gefen$^{1,2}$}

\address{$^1$Center for Theoretical Physics, Sloane Physics Laboratory,
     Yale University, New Haven, Connecticut 06520, USA\\
$^2$ Department of Condensed Matter Physics, The Weizmann 
Institute of Science, Rehovot 76100, Israel }

\date {August 20, 2001}
\maketitle
\begin{abstract}

 We study the fluctuations of an energy level as a function of 
 the number of electrons $m$ added to a 
Coulomb-blockade quantum dot. A microscopic calculation in 
the limit of Koopmans' theorem predicts that the standard 
deviation of these fluctuations behaves as $\sqrt{m}$ in the 
absence of surface charge but is linear in 
$m$ when the effect of a  surface charge in a finite geometry is
included. The microscopic results are compared to a parametric
 random-matrix approach. We estimate the number of electrons
 it takes to scramble the spectrum completely in terms of
 the interaction strength, the dimensionless Thouless 
conductance, and the symmetry class.
\end{abstract}

\pacs{PACS numbers: 73.23.Hk, 05.45+b, 73.21.La, 73.23.-b}

\narrowtext

   The simplest model for describing a quantum dot in the 
Coulomb-blockade regime is the constant 
interaction (CI) model \cite{Kouwenhoven97}, in which the electrons
 occupy single-particle
 levels and the Coulomb interaction is taken as an average 
electrostatic energy that depends only on the number of electrons.
 When the single-particle dynamics  in the dot are chaotic 
(or diffusive), one can apply random matrix theory (RMT) to 
describe the statistical properties of the single-particle 
wave functions and energies within an energy window whose 
width is the Thouless energy. RMT was successful in describing
 the conductance peak height distributions and their sensitivity
 to time-reversal symmetry \cite{JSA92,peaks-exp}. However,
 other measured statistics, most notably the peak spacing
 distribution \cite{spacings-exp}, indicated that it is 
necessary  to include interaction effects beyond the simple
 CI model \cite{Alhassid00}.  The best way to take into account
  interactions while retaining a single-particle framework is the 
Hartree-Fock (HF) approximation, which has been  used to
 explain some of the observed features of the peak spacing
 statistics \cite{HF}. 

  The HF single-particle wave functions and energies are 
calculated self-consistently and therefore can change as
 electrons are added to the dot. In the statistical regime
  (i.e., in a chaotic or diffusive dot),
  this phenomenon is called scrambling. Scrambling was observed
 in the decay of correlations between the $m$-th excited 
state in the dot and the ground-state of a dot with an additional
 $m$ electrons \cite{Stewart97}, and indirectly
  through the saturation of the peak-to-peak 
correlations as a function of temperature 
\cite{Patel98,Alhassid99}. 
 A phenomenological way to
 describe scrambling is to consider a discrete set of Hamiltonians
 (corresponding to the different number of electrons 
in the dot) that are random but have the correct symmetries. 
Such a set is known as a discrete Gaussian process (GP),
 and can be embedded in a continuous GP, i.e., random matrices
  that depend on a
 continuous parameter \cite{AA95}.  This parametric approach leads 
to a nearly Gaussian peak spacing distribution \cite{Vallejos98},
and explains the saturation of the number of correlated peaks 
versus temperature \cite{Alhassid99}. While the parametric 
approach is appealing in its simplicity, it is not clear
 how well it describes features obtained in a  microscopic
 approach.  For example, in 
the parametric approach, the fluctuation standard deviation (FSD) of a typical
 level upon the addition of $m$ electrons into 
the dot is proportional to $m$. One can also evaluate the 
FSD of a level in the microscopic 
approach through the fluctuations of the diagonal interaction
 matrix elements (in the limit where the single-particle 
wave functions do not change with the addition of electrons,
 hence the off-diagonal matrix elements can be neglected). 
 When only the bulk screened Coulomb interaction is accounted, we find that
  the associated FSD behaves like $\sqrt{m}$, unlike the standard parametric
 approach. However, in the finite geometry of a quantum dot, the  level
  fluctuation may be dominated by the variation of the mean-field potential
   arising from surface charge.   We 
shall show that in this case the FSD is linear in $m$ (for $1 \ll m \ll g$,
 where  $g$ is the dimensionless Thouless conductance), in overall agreement
  with parametric RMT.  The main  results of the microscopic approach are
   summarized in Eqs. (\ref{variance-v}), (\ref{covariance-v}) and
    (\ref{variance-m}) in the absence of surface charge and
     Eqs. (\ref{var-surface}), (\ref{covariance-surface}) and 
(\ref{variance-surface}) in the presence of surface charge.
 Extrapolating our results to larger values of $m$,
 we also estimate the dependence of the number of 
added electrons $m_c$ required for complete scrambling on $g$ and
 the interaction strength  both in the absence 
and presence of surface charge (Eqs. (\ref{m-scrambling1}) and
 (\ref{m-scrambling2}), respectively).  

  Scrambling implies variation of both 
eigenstates and energy levels. However, in this work we approach the problem
 from the limit  where the wave functions do not change 
 (Koopmans' theorem \cite{Koopmans}), and we only estimate the  FSD of energy 
 levels. In the parametric approach
 this limit is valid when the change in the parameter upon the addition 
of $m$ electrons is small compared to the mean parametric 
distance  between avoided crossings. 
 Complete scrambling occurs when the energy fluctuations become comparable
  to $\Delta$ (the mean-level spacing in the middle of the band). 

 We first discuss the parametric approach.  The variation
 of the single-particle energies and eigenfunctions (e.g., in a mean-field
 approximation) with the addition of
  electrons into the dot is described by a parametric variation of the
  Hamiltonian \cite{systematic}. We denote by
 $H(x_{\cal N})$ the effective single-particle Hamiltonian of 
the dot with ${\cal N}$ electrons. We assume that the dot is 
either diffusive or ballistic with chaotic dynamics, and that 
 the statistical properties of $H(x_{\cal N})$ are not 
 modified by the interactions. Restricting ourselves to the
 universal regime, i.e., to $\sim g$ levels in the vicinity of 
the Fermi energy \cite{g-vs-N}, we assume that $H(x_{\cal N})$
 belongs to one of the Gaussian ensembles of random matrices whose
 symmetry class $\beta$ is independent of $x_{\cal N}$. The
 sequence of Hamiltonians $H(x_{\cal N})$ forms a discrete GP 
that can be embedded in a continuous GP $H(x)$ \cite{Attias95}.
 A simple GP is given by \cite{AA95,Wilkinson}
$H(x) = \cos x H_1 + \sin x H_2$, 
where $H_1$ and $H_2$ are $N \times N$ uncorrelated random matrices 
chosen from the Gaussian ensemble of symmetry class $\beta$:
 $P(H) \propto e^{-{\beta \over 2 a^2} {\rm Tr} H^2}$. We choose
 $a = (2/N)^{1/2} \zeta$ so that the average level density
 (a semicircle) has a constant band width of
 $2a\sqrt{2N} = 4 \zeta$ and the mean-level spacing in the middle
  of the spectrum  is
$\Delta = \pi a/\sqrt{2 N} = \pi \zeta/N$.  The average distance 
between avoided crossings is given by the inverse of the rms level
 velocity $\delta x_c= \Delta [\overline{(\partial
 \epsilon_\alpha/\partial x)^2}]^{-1/2} = \pi (\beta/2N)^{1/2}$, 
 where $\epsilon_\alpha$ is an energy level. This distance is larger for the
 GUE by a factor of $\sqrt{2}$ compared with the GOE.  Two-point 
parametric correlators become universal when  the parameter $x$
 is measured in units of $\delta x_c$; i.e., as a function of
 a scaled parameter $\bar x \equiv x/\delta x_c$.

  The energy levels $\epsilon_\alpha$ scramble as the parameter 
  $\bar x$ changes:
$\delta \epsilon_\alpha = \epsilon_\alpha(\bar x+ \delta \bar x) 
- \epsilon_\alpha(\bar x)$.  The variance of  $\delta \epsilon_\alpha$
 is estimated in the limit 
$\delta \bar x \ll 1$ to be $\sigma^2(\delta \epsilon_\alpha) = 
\Delta^2 (\delta \bar x)^2$ using first order perturbation theory in 
$\delta \bar x$ (i.e., ignoring the change of the single-particle 
wave function as  $\bar x \to \bar x + \delta \bar x$).

 In the parametric approach it is assumed that $\bar x$ changes
 by $\delta \bar x_1$ upon the addition of one electron into 
the dot (independently of the number of electrons ${\cal N}$).
 Thus for the addition of $m$ electrons 
$\delta \bar x_m = m \delta \bar x_1$, and as long as 
$\delta \bar x_m \ll 1$, we can still use first order perturbation 
theory to find
\begin{equation}\label{var-parametric}
\sigma (\delta \epsilon^{(m)}_\alpha) = 
\Delta m \delta \bar x_1 = m \sigma(\delta \epsilon^{(1)}_\alpha) \;,
\end{equation}
where $\delta \epsilon^{(m)}_\alpha \equiv 
\epsilon^{({\cal N}+m)}_\alpha - \epsilon^{({\cal N})}_\alpha$ 
($\epsilon^{({\cal N})}_\alpha$ is the energy of level 
$\alpha$ in a dot with ${\cal N}$ electrons).

 To relate the parametric approach to a microscopic mean-field
 approach, we describe the effect that adding electrons
  has on a particular single-particle HF level. The Hamiltonian 
of the dot is 
\begin{equation}\label{hamiltonian}
H =  \sum\limits_{ij} h^{(0)}_{ij} a^\dagger_i a_j +
{1\over 4} \sum_{ijkl}
 v^A_{ijkl}a^\dagger_i a^\dagger_j a_l a_k \;,
\end{equation}
where $h^{(0)}$ is the one-body part (accounting for the disorder
 or the chaotic confining potential)  and 
$v^A_{ij;kl} \equiv \langle i j | v| k l \rangle - \langle i j | v| l k
\rangle$  are the antisymmetrized matrix elements of the 
two-body interaction. The HF Hamiltonian in the self-consistent basis 
for ${\cal N}$ electrons is given by
 $h_{\alpha \gamma}^{({\cal N})} =
 \epsilon^{(0)}_{\alpha
\gamma}  + \sum_{\delta=1}^{{\cal N}} v^A_{\alpha \delta;
 \gamma \delta}$ (where the sum is over the lowest ${\cal N}$ 
  occupied levels), and the eigenvalues of $h_{\alpha \gamma}^{({\cal N})}$
 are the HF single-particle energies  
$\epsilon_\alpha^{({\cal N})}$. 
The {\em diagonal} part of the Hamiltonian 
(\ref{hamiltonian}) in the HF basis is
\begin{equation}
H_{\rm diagonal} = \sum\limits_{\alpha} \epsilon^{(0)}_{\alpha \alpha}
 \hat n_\alpha + {1 \over 2} \sum\limits_{\alpha, \gamma} 
v^A_{\alpha \gamma} \hat n_\alpha \hat n_\gamma \;,
 \end{equation}
where $\hat n_\alpha$ is the number operator for single-particle 
state $\alpha$ and $v^A_{\alpha \gamma}
 \equiv v^A_{\alpha \gamma; \alpha \gamma}$. 

In Koopmans' limit, the single-particle wave functions
 do not change when an electron is added to the dot. Consequently, 
the change in the single-particle energy $\epsilon_\alpha$  when
 an electron is added to the ${\cal N}+1$-st level is given by a
 diagonal interaction matrix element
\begin{equation}\label{1-electron}
\delta\epsilon^{(1)}_\alpha \equiv \epsilon^{({\cal N}+1)}_\alpha 
- \epsilon^{({\cal N})}_\alpha \approx v^A_{\alpha {\cal N}+1} \;.
\end{equation}
 
 Assuming the HF wave functions satisfy similar statistics as in
 the non-interacting case, we can estimate the variance of the
 matrix element in (\ref{1-electron}). The two-body interaction used in
  the HF approximation is often
 an effective interaction, e.g., an RPA screened interaction 
\cite{Blanter97}. Alternatively, one can employ a 
 short-range dressed interaction \cite{Altshuler97} 
$v(\bbox r - \bbox r') = \lambda \Delta V \delta 
(\bbox r - \bbox r')$, where $\lambda$ is the interaction 
strength and $V$ is the system's volume. A diagonal interaction
 matrix element is then given by 
$v_{\alpha \gamma}= \lambda \Delta V \int d\bbox r 
\psi^\ast_\alpha(\bbox r) \psi^\ast_\gamma(\bbox r)
\psi_\alpha(\bbox r) \psi_\gamma(\bbox r)$, and its variance
\begin{eqnarray}\label{variance-diag}
\sigma^2(v_{\alpha \gamma}) & &  =  \lambda^2 \Delta^2 V^2 
\int d \bbox r_1 \int d \bbox r_2 \left[ \left\langle 
\psi^\ast_\alpha(\bbox r_1) \psi^\ast_\gamma(\bbox r_1)
\psi_\alpha(\bbox r_1) \psi_\gamma(\bbox r_1)
\psi_\alpha(\bbox r_2) \psi_\gamma(\bbox r_2)
\psi^\ast_\alpha(\bbox r_2) \psi^\ast_\gamma(\bbox r_2)
 \right\rangle \right. \nonumber \\
& & - \left. \left\langle \psi^\ast_\alpha(\bbox r_1) 
\psi^\ast_\gamma(\bbox r_1)
\psi_\alpha(\bbox r_1) \psi_\gamma(\bbox r_1) \right\rangle 
\left\langle \psi_\alpha(\bbox r_2) \psi_\gamma(\bbox r_2)
\psi^\ast_\alpha(\bbox r_2) \psi^\ast_\gamma(\bbox r_2) 
\right\rangle  \right] \;.
\end{eqnarray}
 Only the connected part of the ensemble average over 
the product of eight wave functions contributes to the r.h.s.
 of (\ref{variance-diag}). 
 In the following we restrict our 
calculations to the GUE symmetry. All possible pairwise and 
quadruplet-wise contractions  should
 be taken into account, and they can be regrouped to give 
\begin{eqnarray}\label{quadruplets}
\sigma^2(v_{\alpha \gamma})& & =  \lambda^2 \Delta^2 V^2 \int 
d \bbox r_1 \int d \bbox r_2  \left\{ \left[  \left\langle 
\psi^\ast_\alpha(\bbox r_1) \psi_\alpha(\bbox r_1)
\psi_\alpha(\bbox r_2) \psi^\ast_\alpha(\bbox r_2) \right\rangle
\left\langle \psi^\ast_\gamma(\bbox r_1) \psi_\gamma(\bbox r_1)
\psi_\gamma(\bbox r_2) \psi^\ast_\gamma(\bbox r_2) \right\rangle 
 -{1\over V^4} \right] \right. \nonumber \\
    +  & & \left[  \left\langle 
\psi^\ast_\alpha(\bbox r_1) \psi_\alpha(\bbox r_1)
\psi_\gamma(\bbox r_2) \psi^\ast_\gamma(\bbox r_2) \right\rangle
\left\langle \psi^\ast_\gamma(\bbox r_1) \psi_\gamma(\bbox r_1)
\psi_\alpha(\bbox r_2) \psi^\ast_\alpha(\bbox r_2) \right\rangle 
-{1\over V^4} \right]  \nonumber \\
 + & & \left.  \left\langle 
\psi^\ast_\alpha(\bbox r_1) \psi_\alpha(\bbox r_2)
\psi_\gamma(\bbox r_1) \psi^\ast_\gamma(\bbox r_2) \right\rangle
\left\langle \psi^\ast_\gamma(\bbox r_1) \psi_\gamma(\bbox r_2)
\psi_\alpha(\bbox r_1) \psi^\ast_\alpha(\bbox r_2) \right\rangle  
  \right\} \;,
\end{eqnarray}
where the $1/V^4$ factors are subtracted to avoid double
 counting of pairwise contractions (here $\langle\ldots\rangle$ 
 denotes an ensemble average and not just the connected part).

 Wave-function correlations appearing in Eq. (\ref{quadruplets})
 were calculated in Ref. \cite{Blanter97',Mirlin00}. Including
 contributions of order $1/g^2$ and assuming that the distance 
 between levels $\alpha$ and $\gamma$  is less than the Thouless
  energy, we have
\begin{eqnarray}\label{corr}
\sigma^2(v_{\alpha \gamma})& & = {\lambda^2 \Delta^2\over V^2}
 \int d \bbox r_1 \int d \bbox r_2  \left\{ \left[ 1+k_d(r)
(1+\Pi(\bbox r_1, \bbox r_2)) + \Pi(\bbox r_1, \bbox r_2) +
 b(\bbox r_1, \bbox r_2)\right]^2 -1 \right. \nonumber \\
    +  & & \left. \left[ 1 + k_d(r) \Pi(\bbox r_1, \bbox r_2) + 
 c(\bbox r_1, \bbox r_2) \right]^2  -1  
 + \left[ k_d(r) + \Pi(\bbox r_1,\bbox r_2) 
\right]^2  \right\} \;.
\end{eqnarray}  
The function $k_d(r)$ ($\bbox r = \bbox r_1 - \bbox r_2$) is given
 by  $k_d(r) \equiv (\pi \nu)^2\langle  
{\rm Im} G^R(\bbox r)\rangle^2$ (where $G^R$ is the retarded
 Green function and $\nu$ the electron density of states per unit
 volume) and describes the short-range correlations.  In 2D 
$k_d(r) = \exp(-r/\ell)J^2_0(k_F r)$ (where $\ell$ is the mean
 free path and $J_0$ is a Bessel function),
 while in 3D $k_d(r) = \exp(-r/\ell) (\sin k_F r/ k_F r)^2$. 
 $\Pi(\bbox r_1, \bbox r_2) = 
(\pi \nu)^{-1}\sum_{\bbox Q \neq 0} \phi_{\bbox Q}(\bbox r_1)
 \phi_{\bbox Q}(\bbox r_2) /D \bbox Q^2$ is the long-range 
diffuson propagator where $\phi_{\bbox Q}$ is the eigenfunction of
 the diffusion operator corresponding to the eigenvalue 
$D \bbox Q^2$ ($D$ is the diffusion constant).  
 The functions $b,c$ in Eq.~(\ref{corr}) are 
of order $1/g^2$ ($c \approx \Pi^2(\bbox r_1, \bbox r_2)/2$ for 
$ \ell \ll r \ll L$ \cite{Mirlin00} where $L$ is the linear size of the dot). 
The main contributions to (\ref{corr}) arise from 
the long-range terms $\Pi^2$ ($\int d \bbox r_1 
\Pi(\bbox r_1, \bbox r_2) =0$ and does not contribute). 
Using the
 normalization conditions $\int d \bbox r_1 \int d \bbox r_2  \left\langle 
\psi^\ast_\alpha(\bbox r_1) \psi_\alpha(\bbox r_1)
\psi_\alpha(\bbox r_2) \psi^\ast_\alpha(\bbox r_2) \right\rangle = 1$ and 
$\int d \bbox r_1 \int d \bbox r_2  \left\langle 
\psi^\ast_\alpha(\bbox r_1) \psi_\alpha(\bbox r_1)
\psi_\gamma(\bbox r_2) \psi^\ast_\gamma(\bbox r_2) \right\rangle = 1$, it follows 
 that \cite{Mirlin00,Baranger}
 $\int d\bbox r_1 \int d\bbox r_2 \left[ k_d(r)
 \left(1+ \Pi(\bbox r_1,\bbox r_2)\right) + b(\bbox r_1, \bbox r_2)\right] = 0$ and
$\int d\bbox r_1 \int d\bbox r_2 \left[ 1 + k_d(r)\Pi(\bbox r_1, \bbox r_2)
 + c(\bbox r_1, \bbox r_2)\right] = 0$, respectively. 
Thus the functions $b$ and $c$ do not contribute  to order $1/g^2$ and
\begin{equation}\label{variance-v}
\sigma^2(v_{\alpha \gamma}) = 2 \kappa \left(\lambda 
\Delta / g \right)^2 \;,
\end{equation}
 where $g=2\pi \nu D L^{d-2}$ is the dimensionless Thouless
 conductance. For a cube in $d$ dimensions 
$\kappa ={4 \over \pi^4} \sum_{{\bbox n} \neq 0} 
{1/ |\bbox n|^4}$, where
the summation is over $\bbox n =(n_1,\ldots,n_d)$
 with $n_i$ non-negative integers
[${\bbox n} \neq (0,\ldots,0)$]. 

  We next consider the addition of $m$ electrons into the dot. 
 Extending Koopmans' limit to $m$ electrons, i.e., 
assuming the single-particle wave functions do not change 
with the addition of $m$ electrons, we can express the change 
in a single-particle level $\alpha$ as
\begin{equation}\label{m-electrons}
\delta \epsilon^{(m)}_\alpha \equiv \epsilon^{({\cal N}+m)}_\alpha
 - \epsilon^{({\cal N})}_\alpha \approx \sum_{i=1}^{m}
 v^A_{\alpha {\cal N} +i} \;.
\end{equation}
For $m <g$, the variance of (\ref{m-electrons}) is given by 
$\sigma^2(\delta \epsilon^{(m)}_\alpha) = m \sigma^2(v_{\alpha \gamma}) 
+ m(m-1) 
(\overline{v_{\alpha \gamma}v_{\alpha \delta}} - 
\overline{v_{\alpha \gamma}}\; \overline{ v_{\alpha \delta}})$ 
 (since the variances and covariances are independent of the 
particular wave functions). In  analogy to Eq. (\ref{quadruplets}), 
the covariance of $v_{\alpha \gamma}$ and $v_{\alpha \delta}$ is given by
\begin{eqnarray}\label{quadruplets1}
\overline{v_{\alpha \gamma}v_{\alpha \delta}} & - & 
\overline {v_{\alpha \gamma}}\; \overline{ v_{\alpha \delta}} = \nonumber \\
  \lambda^2 &\Delta^2 & V^2  \int 
d \bbox r_1 \int d \bbox r_2  \left\{ \left[  \left\langle 
\psi^\ast_\alpha(\bbox r_1) \psi_\alpha(\bbox r_1)
\psi_\alpha(\bbox r_2) \psi^\ast_\alpha(\bbox r_2) \right\rangle
\left\langle \psi^\ast_\gamma(\bbox r_1) \psi_\gamma(\bbox r_1)
\psi_\delta(\bbox r_2) \psi^\ast_\delta(\bbox r_2) \right\rangle 
 -{1\over V^4} \right] \right. \nonumber \\
   & & +  \left. \left[  \left\langle 
\psi^\ast_\alpha(\bbox r_1) \psi_\alpha(\bbox r_1)
\psi_\delta(\bbox r_2) \psi^\ast_\delta(\bbox r_2) \right\rangle
\left\langle \psi^\ast_\gamma(\bbox r_1) \psi_\gamma(\bbox r_1)
\psi_\alpha(\bbox r_2) \psi^\ast_\alpha(\bbox r_2) \right\rangle 
-{1\over V^4} \right] 
  \right\} \;,
\end{eqnarray}
and using the known wave-function correlations 
\begin{eqnarray}\label{corr1}
\overline{v_{\alpha \gamma}v_{\alpha \delta}} & - & 
\overline {v_{\alpha \gamma}}\; \overline{ v_{\alpha \delta}}  =  {\lambda^2 
\Delta^2\over V^2} \int d \bbox r_1 \int d \bbox r_2  
\left\{ \left[ 1+k_d(r)(1+\Pi(\bbox r_1, \bbox r_2)) + 
\Pi(\bbox r_1, \bbox r_2) + b(\bbox r_1, \bbox r_2)\right]
 \right. \nonumber \\ 
 & \times & \left. \left[ 1 + k_d(r) \Pi(\bbox r_1, \bbox r_2) + 
c(\bbox r_1, \bbox r_2)\right] -1 +  \left[ 1 + k_d(r) 
\Pi(\bbox r_1, \bbox r_2) + 
c(\bbox r_1, \bbox r_2) \right]^2  -1   \right\} \;.
\end{eqnarray}
Terms  such as $k_d^2\Pi$, $k_d\Pi^2$ and  $k_d c$ in Eq. (\ref{corr1})
 contribute $\sim (\ell/L)^2 \lambda^2 \Delta^2/g^3$ and behave like 
 $\sim \lambda^2 \Delta^2/g^3$ in the 
ballistic limit $\ell \sim L$. Other terms contribute $\sim 1/g^4$. 
Thus to leading order in $1/g^2$
\begin{equation}\label{covariance-v}
\overline{v_{\alpha \gamma}v_{\alpha \delta}} - 
\overline {v_{\alpha \gamma}}\;\overline{ v_{\alpha \delta}} \approx 0 \;.
\end{equation}
 Using Eqs. (\ref{m-electrons}), (\ref{variance-v}) and (\ref{covariance-v})
  we find \cite{covariance}
\begin{equation}\label{variance-m}
\sigma^2(\delta\epsilon^{(m)}_\alpha) =  
m \sigma^2(\delta\epsilon^{(1)}_\alpha) = 2 m \kappa (\lambda \Delta/g)^2
 \;.
\end{equation}
  We conclude that the 
statistical fluctuations accumulated by sequentially adding
 electrons into the dot are added independently, namely, the FSD of an 
 energy level behaves like $\sqrt{m}$, 
 contrary to the parametric approach where it is linear in $m$ 
 (see Eq. (\ref{var-parametric})). 

 A complete scrambling of the spectrum corresponds to a number of electrons 
$m_c$ for which  
$\sigma(\delta\epsilon^{(m_c)}_\alpha) \sim \Delta$. Using (\ref{variance-m})
we find
\begin{equation}\label{m-scrambling1}
 m_c \sim \left({g \over \lambda}\right)^2 \;.
\end{equation}
We note that since we have neglected wave-function
 scrambling due to off-diagonal interaction matrix elements, our  estimate
in Eq. (\ref{m-scrambling1}) should 
only be regarded as an upper bound.

 Next we consider the effect due to excess negative charge on the 
boundaries of the dot. The effective interaction due to such charge is 
\cite{Blanter97}
$v(\bbox r, \bbox r') = V_\kappa(\bbox r) + V_\kappa(\bbox r')$ 
where $V_\kappa(\bbox r)$  describes
the variation of the mean-field potential upon the addition of one electron. 
For a disk 
of radius $R$ in 2D we have $V_\kappa(\bbox r) = 
- (e^2/2 \kappa R)(R^2 -r^2)^{-1/2}$ where $\kappa$
is the inverse screening length. The related interaction  matrix element is now  
$v_{\alpha \gamma} = \int d\bbox r |\psi_\alpha(\bbox r)|^2 V_\kappa(\bbox r) +
 \int d\bbox r |\psi_\gamma(\bbox r)|^2 V_\kappa(\bbox r)$. 
Using the wavefunction correlations to order $1/g^2$, its variance is
\begin{eqnarray}\label{surface}
\sigma^2 (v_{\alpha \gamma}) &  =  & { 2 \over  V^2}
 \int d \bbox r_1 \int d \bbox r_2  \left[
\left(V^2 \langle |\psi_\alpha(\bbox r_1)|^2 |\psi_\alpha(\bbox r_2)|^2\rangle
 -1 \right) + \left(V^2 \langle |\psi_\alpha(\bbox r_1)|^2 
|\psi_\beta(\bbox r_2)|^2\rangle  -1 \right)\right] \nonumber \\
 && \hspace*{2 cm} \times V_\kappa(\bbox r_1) V_\kappa(\bbox r_2) 
     = 
 { 4\over \beta  V^2}
 \int d \bbox r_1 \int d \bbox r_2   \{k_d(r)
[1+ (\beta +2)\Pi(\bbox r_1, \bbox r_2)/\beta]  \nonumber 
\\ && \hspace*{5.5 cm} + \Pi(\bbox r_1,\bbox r_2) 
 + b(\bbox r_1, \bbox r_2) +  c(\bbox r_1, \bbox r_2) \} 
V_\kappa(\bbox r_1) V_\kappa(\bbox r_2) \;.
\end{eqnarray}
The leading order contribution to (\ref{surface}) comes from 
$\Pi(\bbox r_1,\bbox r_2)$. Using the expansion of 
$\Pi(\bbox r_1,\bbox r_2)$ 
in terms of the diffusion modes $\phi_{\bbox Q}$, the respective
 contribution is
\begin{equation}\label{var-surface}
 \sigma^2(v_{\alpha \gamma}) = {4 \over \beta V^2} {1 \over \pi \nu} 
\sum\limits_{\bbox Q \neq 0} 
{\left[\int d \bbox r \phi_{\bbox Q}(\bbox r) V_\kappa(\bbox r) 
\right]^2 \over D\bbox Q^2} = a{\lambda^2 \Delta^2 \over \beta   g} \;.
\end{equation}
 For a disk in 2D the coefficient $a = (2 \pi)^{-1}\sum_{\bbox q}
  \left[\int d\bbox r \phi_{\bbox q}(\bbox r)
(1-r^2)^{-1/2}\right]^2/q^2$, where  $\phi_{\bbox q}$ are 
the diffusion modes in a disk of radius $R=1$ ($\bbox Q = R^{-1} \bbox q$).
 The short-range term $k_d$ in (\ref{surface}) contributes 
a term $\sim (\ell/L)^2 (\Delta^2 /\beta g)$ that for a diffusive dot 
is parametrically small compared with (\ref{var-surface}), but in the 
ballistic limit its contribution becomes comparable to (\ref{var-surface}). 
 The surface charge contribution to the variance (\ref{var-surface}) 
 dominates the bulk contribution (\ref{variance-v}). The bulk-surface 
  cross-correlation term is given by $(2 \lambda \Delta/V^2) 
   \int d \bbox r_1 \int d \bbox r_2 
\left[ V^2\langle|\psi_\alpha(\bbox r_1)|^2 |\psi_\alpha(\bbox r_2)|^2\rangle
 -1 \right] V_\kappa(\bbox r_2)$. The dominating contribution due to $k_d$ 
  ($\sim (\ell/L)^2  \Delta^2/g$) is suppressed in a diffusive dot but is
   comparable to (\ref{var-surface}) in the ballistic limit.  
    
 Similarly, the covariance of  $v_{\alpha \gamma}$ and $v_{\alpha \delta}$
(due to surface charge) is evaluated to be 
 \begin{eqnarray}\label{surface'}
& & \overline{v_{\alpha \gamma} v_{\alpha \delta}}   - 
\overline {v_{\alpha \gamma}}\;\overline{ v_{\alpha \delta}}  \nonumber \\
   & & =  {1 \over  V^2}
 \int d \bbox r_1 \int d \bbox r_2  \left[
\left(V^2 \langle |\psi_\alpha(\bbox r_1)|^2 |\psi_\alpha(\bbox r_2)|^2\rangle
 -1 \right) + 3\left(V^2 \langle |\psi_\alpha(\bbox r_1)|^2 
|\psi_\beta(\bbox r_2)|^2\rangle  -1 \right)\right] 
   V_\kappa(\bbox r_1) V_\kappa(\bbox r_2)   \nonumber \\
 &  & = { 2 \over \beta  V^2}
 \int d \bbox r_1 \int d \bbox r_2  \left[ k_d(r)
\left(1+ {3 \beta +2\over \beta}\Pi(\bbox r_1, \bbox r_2)\right) + 
\Pi(\bbox r_1, \bbox r_2) +
 b(\bbox r_1, \bbox r_2) + 3c(\bbox r_1, \bbox r_2)
 \right] \\ \nonumber & & \hspace*{12 cm} \times V_\kappa(\bbox r_1) 
V_\kappa(\bbox r_2) \;.
\end{eqnarray}
The leading order contributions for both a diffusive and ballistic dots are 
similar to the ones contributing to the variance, and comparing 
(\ref{surface'}) with (\ref{surface}), we find
\begin{equation}\label{covariance-surface}
\overline{v_{\alpha \gamma}v_{\alpha \delta}} - 
\overline {v_{\alpha \gamma}}\;\overline{ v_{\alpha \delta}} \approx 
{1\over 2}\sigma^2(v_{\alpha \gamma}) \;.
\end{equation}  
We therefore obtain the following relation
\begin{equation}\label{variance-surface}
\sigma^2(\delta\epsilon_\alpha^{(m)}) = 
{1 \over 2} m(m+1) \sigma^2(\delta \epsilon_\alpha^{(1)}) 
= {1\over 2}m(m+1) a { \Delta^2 \over \beta g} \;.
\end{equation}

 Thus, for $ m \gg 1$, the FSD of a level is linear in the number of added
  electrons $m$, in agreement with the result (\ref{var-parametric})
 of the parametric approach.  The scrambling parameter $\delta \bar x_1$ 
 can be determined by comparing the variance (\ref{var-parametric}) 
 in the parametric approach with  the corresponding  variance
  (\ref{variance-surface}) in the microscopic approach 
\begin{equation}\label{delta-x}
\delta \bar x_1 =  \left({a \over 2 \beta g}\right)^{1/2} \;. 
\end{equation}

A  complete scrambling of the spectrum corresponds to a parametric 
change of one avoided crossing, i.e., $\delta \bar x_m \sim 1$.
 Using (\ref{delta-x}), we find
\begin{equation}\label{m-scrambling2}
m_c \sim {1 \over \delta \bar x_1} \sim   (\beta g)^{1/2} \;.
\end{equation}
When an interaction scale $\lambda$ is introduced (e.g., screening due to
 external gates), we find $m_c \sim (\beta g)^{1/2}/\lambda$. 
 
In a recent experiment \cite{Maurer99} the standard deviation 
$\sigma_p(m) \equiv \sigma(V_{{\cal N} +m} - V_{\cal N})$, 
where $V_{\cal N}$ is the position in gate voltage for the
 ${\cal N}$-th peak, was measured as a function of the separation
 $m$ in peak number. In Koopmans' limit 
$V_{{\cal N} +m} - V_{\cal N} \approx 
\epsilon^{({\cal N} +m)}_{{\cal N} +m } - 
\epsilon^{({\cal N})}_{\cal N} = \Delta_m^{({\cal N} + m)}+
 \delta\epsilon^{(m)}_{\cal N}$, where $\Delta_m^{({\cal N} + m)}
 \equiv \epsilon^{({\cal N} +m)}_{{\cal N} +m } - 
\epsilon^{({\cal N} +m)}_{{\cal N}}$ is the distance between levels 
separated by $m$ consecutive spacings in a dot with a fixed number 
of electrons (${\cal N} +m$). The latter quantity 
 is unrelated to scrambling and its fluctuation for $m<g$ is 
completely determined from RMT:  
$\sigma(\Delta_m^{({\cal N} + m)}) \propto \Delta \ln m$.
 In principle one can extract $\sigma(\delta\epsilon^{(m)}_{\cal N})$
  from the measured $\sigma_p(m)$. In practice this is difficult
 to do with the available experimental statistics since for $m < m_c$,
  $\sigma(\delta\epsilon^{(m)}_{\cal N})$ is smaller than
   $\sigma(\Delta_m^{({\cal N} + m)})$, and 
$\sigma_p(m)$ is dominated by the $\ln m$ term of RMT. 
Nevertheless, $\sigma_p(m)$, when plotted versus $\ln m$, appears to
 follow a straight-line behavior with a small concave deviation 
 \cite{Marcus}, 
in qualitative agreement with an additional contribution due to
 scrambling.  

In conclusion, we have studied spectral scrambling when several
 electrons are added to a Coulomb-blockade quantum dot.  
In the absence of surface charge the fluctuation standard deviation of
 a level upon the addition of $m$ electrons behaves like $\sqrt{m}$, 
 but when surface charge is present this standard deviation  
is linear in $m$. The latter result is the one obtained in the
 conventional parametric random matrix approach.  
  Our analysis was carried out within the framework 
of the HF-Koopmans picture and was limited to energy fluctuations.
 A microscopic description of wave functions scrambling 
is an outstanding problem.

We acknowledge H. Baranger, C. M. Marcus and A. D. Mirlin for useful
 discussions.  We also thank S. Altman
 for a Yale-Weizmann cooperation fellowship to support the visit of Y.G. to 
 Yale, where part of this work was completed.
  This work was supported in part by the Department of Energy grant
No. DE-FG-0291-ER-40608, by the U.S.-Israel Binational Science
 Foundation, by the DIP Foundation, by the ISF of the Israel Academy
 of Science and Humanities, and the Minerva Foundation.


\begin{references}
\bibitem{systematic} We subtract the systematic dependence of 
$\{\epsilon_\lambda \}$ on ${\cal N}$.
\bibitem{g-vs-N} We note that the relation between the rank $N$ 
of the random matrix and $g$ is rather ambiguous. In a single-particle
 theory $g \sim N$. However, if one
 is interested in the fluctuations of the interaction 
matrix elements, then the correct identification is $g \sim N^{1/2}$.
\bibitem{covariance} The covariance (\ref{covariance-v}) cannot
 always be ignored. The number of such terms in 
 $\sigma^2(\delta\epsilon^{(m)}_\alpha)$ is $\sim m^2$ and for a ballistic dot
 with $m > g$ their contribution might dominate. However, for $m> g$
the distance between some of the levels involved is larger than the
Thouless energy and new estimates are required. Details will be
 published elsewhere.
\end{references}
\end{document}